\documentclass[a4paper]{article}
\usepackage[english]{babel}
\usepackage{listings}
\usepackage{fancyhdr}
\usepackage{graphicx, color}
\usepackage{hyperref}
\usepackage{lineno, blindtext}
\usepackage[section]{placeins}
\usepackage{algorithm}
\usepackage{algorithmicx}
\usepackage{algpseudocode}
\usepackage{comment}
\usepackage{amssymb}
\usepackage{amstext}
\usepackage{amsthm}
\usepackage{amsmath, amsfonts, mathtools}
\usepackage{pdfpages}
\usepackage[paper=a4paper,margin=3.5cm, bmargin=2.5cm]{geometry}
\usepackage{fancyhdr, titling, setspace, anyfontsize}
\usepackage{fancyvrb}
\usepackage{enumerate}
\usepackage{subcaption}
\usepackage{float}
\usepackage{todonotes}
\usepackage{titlesec}
\usepackage[margin=1.5cm]{caption}

\usepackage{csquotes}
\usepackage[style=numeric, backend=biber, sorting=none]{biblatex}

\bibliography{references}

\algdef{SE}[DOWHILE]{Do}{DoWhile}{\algorithmicdo}[1]{\algorithmicwhile\ #1}

\def\titletext{BLAS-like Interface for Binary Tensor Contractions}
\def\name{Niklas Hörnblad}

\title{
    \includegraphics[width=1\textwidth]{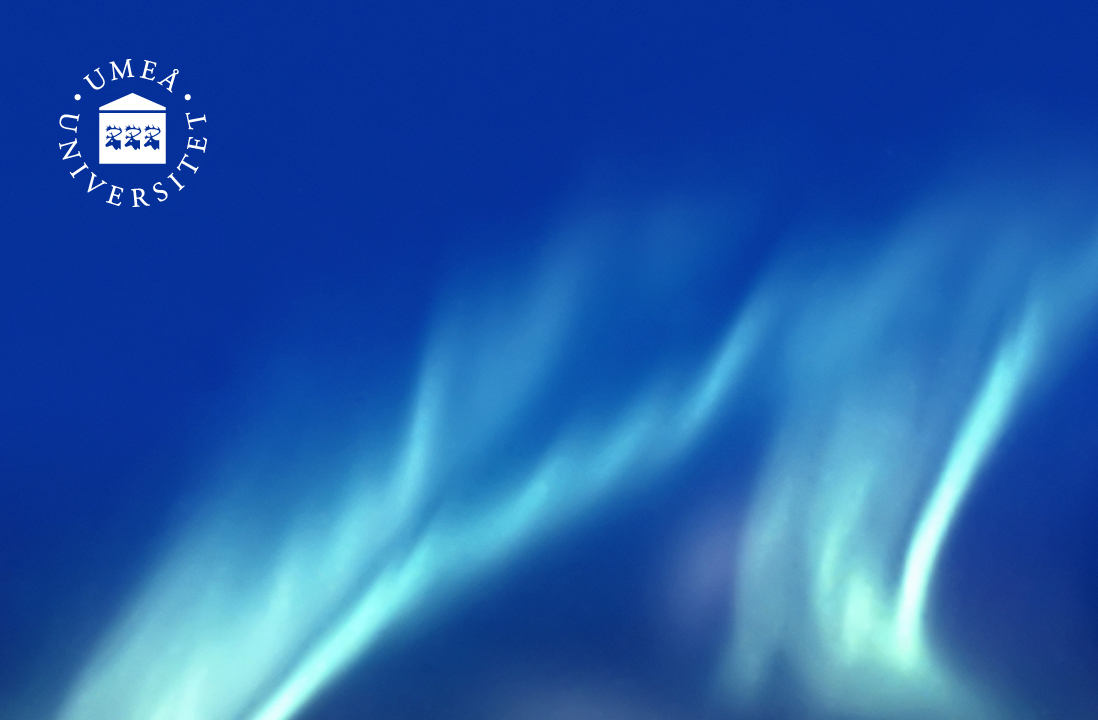} \\
    \vspace{0.7cm}
    Master Thesis Report\\
    \vspace{0.6cm}
    \LARGE \titletext\\
    \vspace{0.8cm}
}

\author{
    \name\\
}
\date{\today}

\begin{document}

\newgeometry{margin=1.1cm, lmargin=1.4cm, top=0cm}
\pretitle{\begin{flushleft} \fontsize{30}{34}\selectfont}
\posttitle{\end{flushleft}}
\preauthor{\begin{flushleft} \linespread{1.4} \large}
\postauthor{\end{flushleft}}
\predate{}
\postdate{}

\maketitle
Examiner: Henrik Björklund\\
Supervisor: Paolo Bientinesi
\vfill
\begin{flushleft}
    \small
    Fall 2024 \\
    Master Thesis, 30 ECTS\\
    Master of Science in Computing Science and Engineering, 300 ECTS
\end{flushleft}

\thispagestyle{empty}
\restoregeometry

\lfoot{\footnotesize{\name}}
\rfoot{\footnotesize{\today}}
\lhead{\sc\footnotesize\titletext}
\rhead{\nouppercase{\sc\footnotesize\leftmark}}
\pagestyle{fancy}
\renewcommand{\headrulewidth}{0.2pt}
\renewcommand{\footrulewidth}{0.2pt}
\pagenumbering{gobble}

\section*{Abstract}
\label{sec:abstract}
In the world of linear algebra computation, a well-established standard exists called BLAS(Basic Linear Algebra Subprograms). This standard has been crucial for the development of software using linear algebra operations. Its benefits include portability with efficiency and mitigation of suboptimal re-implementations of linear algebra operations. Multilinear algebra is an extension of linear algebra in which the central objects are tensors, which are generalizations of vectors and matrices. Though tensor operations are becoming more common, they do not have a standard like BLAS. Such standardization would be beneficial and decrease the now-visible replication of work, as many libraries nowadays use their own implementations. This master thesis aims to work towards such a standard by discovering whether or not a BLAS-like interface is possible for the operation \textit{binary tensor contraction}. To answer this, an interface has been developed in the programming language C together with an implementation and tested to see if it would be sufficient. The interface developed is:
\\\\
\texttt{xGETT(RANKA, EXTA, \hphantom{ }INCA, \hphantom{ }A,\\
\null\qquad \quad RANKB, EXTB, \hphantom{ }INCB, \hphantom{ }B,\\
\null\qquad \quad CONTS, CONTA, CONTB, PERM,\\
\null\qquad \qquad \qquad \qquad \qquad \hphantom{ }INCC, \hphantom{ }C)}
\\\\
with the implementation and tests, it has been deemed sufficient as a BLAS-like interface for \textit{binary tensor contractions} and possible to use in a BLAS-like standardization for tensor operations.

\newpage

\pagenumbering{roman}
\tableofcontents

\newpage
\pagenumbering{arabic}

\setlength{\parindent}{0pt}
\setlength{\parskip}{10pt}

\section{Introduction}
\label{sec:introduction}
    Linear algebra is a field of mathematics concerning linear equations. It is used for signal processing, statistics, machine learning, and data science. Central objects in the field are vectors and matrices, which essentially are one-way and two-way arrays, respectively. In the last few decades, a major advancement has gone beyond the linear world to the multilinear world. This advancement has created an extension to linear algebra called multilinear algebra \autocite{evert2023tensors}. Instead of vectors and matrices, the central object in multilinear algebra is the tensor, which is a generalization of vectors and matrices. A tensor is essentially an n-way array. 
    
    Multilinear algebra has several uses in today's world of computing science. It is used in fields such as machine learning and modern data science\autocite{tensor_machine_learning}, quantum chemistry and physics\autocite{tensor_quantum_chemistry}, signal and image processing\autocite{tensor_image_processing}, chemometrics\autocite{tensor_chemomectrics}, and biochemistry \autocite{tensor_biochemistry}. 
    
    Looking at the landscape of software for tensor computations, numerous libraries for different languages and with different support for tensor operations can be found \autocite{psarras2021landscape}. All these libraries have different implementations for these operations. Those multiple implementations are a massive replication of effort.

    BLAS(Basic Linear Algebra Subprograms) is an interface standardization \autocite{blas_1} \autocite{blas_2} \autocite{blas_3}. It provides a building block that lets libraries access linear algebra operations through the interfaces and use already-existing implementations that can be highly optimized. This has prevented large-scale duplicate work and propelled the usage of basic linear algebra forward.

    This kind of standardization does not exist for tensor operations, though it would be needed and could prevent further duplicate work on the tensor front. This master thesis aims to work towards a BLAS-like standardization for tensors. The master thesis is limited to answering whether an interface in the style of BLAS is sufficient to express an arbitrary \textit{binary tensor contraction} between two tensors of arbitrary dimensions, including support for subtensors.

    To answer this, an interface in a basic BLAS-like style has been developed with an associated example implementation of the operation. To ensure the correctness and full functionality of the interface and its implementation, several tests have been developed.

\section{Background}
\label{sec:background}
    This section explains the underlying mathematical theory needed to understand the operation at the center of this master thesis. It will also explain BLAS, an interface standardization which also is a central part of this master thesis, as it is for vector and matrix operations what this paper works towards achieving for tensor operations.
    \subsection{Tensor}
    \label{sec:background:tensor}
        Tensors are multidimensional mathematical objects that generalize the concepts of scalars, vectors, and matrices by using an arbitrary number of indices(dimensions) \autocite{mathworld_tensor}. A tensor of rank zero is a scalar, a tensor of rank one is a vector and a tensor of rank two would be a matrix. Tensor notation uses sub-scripts and super-scripts to represent the indices of the tensor. The notation ${A^{i j k}}_{l m n}$ denotes a tensor. This is a tensor of rank six, where $i,j,k,l,m,n$ represent the indices of the tensor. Indices written with superscripts can be interpreted as general "row modes" and indices written with subscripts as general "column modes" \autocite{di2013efficient}. These types of indices have compatibility significance in the mathematical notation of tensor operations, similar to how in matrix multiplication, one matrix's columns need to line up with the other matrix's rows. Their significance in binary tensor contractions is explained in \autoref{sec:background:binary_tensor_contraction}.
        
        In computing science, these types of indices are seldom differentiated, the same way column and row vectors are seldom differentiated. But for context in physics, indices written with superscripts are \textit{covariant} indices, and indices written with subscripts are \textit{contravariant}. These are two ways of representing a vector in a coordinate system with a basis. The \textit{contravariant} representation is the more familiar way of representing a vector by parallel projection onto the basis vectors \autocite{Kusse2010-nz}. The \textit{covariant} representation is probably less familiar. It is represented by a perpendicular projection onto the basis vectors. 
        
        The \textit{contravariant} representation looks like $\vec{a}=\sum_{i=1}^{n}a^{i} \times \vec{e_{i}}$, where $\vec{e_{i}}$ are the basis vectors and $a^{i}$ are the magnitude of each basis vector needed to reach the point $\vec{a}$ represents. In this representation, a change in the basis vector will have a contrary effect on $a^{i}$. If a basis vector $\vec{e_{x}}$ doubles, $a^{x}$ will be halved, and if $\vec{e_{x}}$ is halved, $a^{x}$ will double.

        The \textit{covariant} representation looks like $\vec{b}=\sum_{i=1}^{n}b^{i} \times \vec{e_{i}}$, where $\vec{e_{i}}$ are the basis vectors and $b^{i}$ are the lengths of perpendicular projection of the \textit{contravariant} representation of $\vec{b}$ onto the basis vector. Such a projection onto $\vec{e_{x}}$ is calculated by $proj_{\vec{e_{x}}}\vec{b}=\frac{\vec{b} \cdot \vec{e_{x}}}{||\vec{e_{x}}||^{2}} \times \vec{e_{x}}$ \autocite{Anton2013-zm}. Because parallel projections do not change depending on the length of the object projected onto, $b^{i}$ will not change if the length of the basis vectors changes. Therefore, a change in basis will have a similar effect on the \textit{covariant} representation of $\vec{b}$.

\begin{figure}
            \centering
            \includegraphics[scale=0.20]{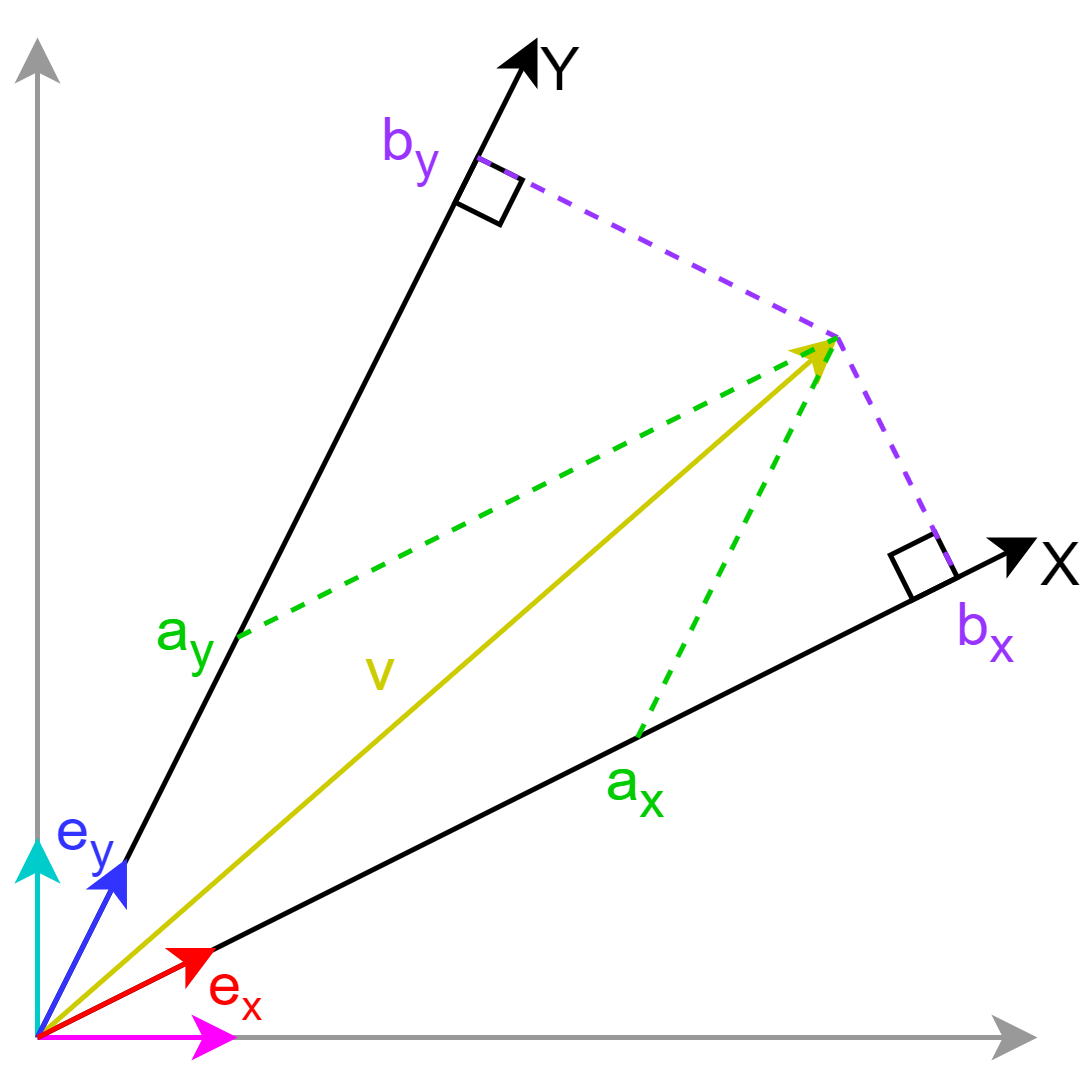}
            \caption{The difference between the \textit{contravariant} and \textit{covariant} ways of representing the yellow vector $v$. $e_y$ and $e_x$ are basis vectors, $a_y$ and $a_x$ are the \textit{contravariant} ways of representing the vector, and $b_y$ and $b_x$ are the covariant ways of representing the vector. The pink and turquoise arrows are more recognizable as (1, 0) and (0, 1) basis vectors.}
            \label{fig:projections}
        \end{figure}

        \autoref{fig:projections} is a visualization of the \textit{covariant} and \textit{contravariant} representations of a vector in 2D space.

    \subsection{Binary Tensor Contraction}
    \label{sec:background:binary_tensor_contraction}
\begin{figure}
            \centering
            \includegraphics[scale=0.35]{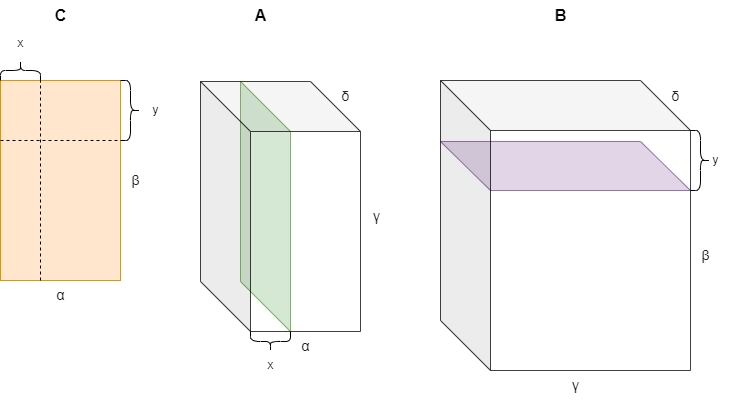}
            \caption{Visualization of the sub-tensors used to calculate the value at position $x$, $y$ in the output tensor.}
            \label{fig:part_tensor_contraction}
        \end{figure}

        A binary tensor contraction is a generalization of the dot product and the matrix product \autocite{di2013efficient}. Instead of being limited to vectors and matrices, it applies to tensors. The operation uses two input tensors to create a third output tensor, which is a combination of the inputs. A binary tensor contraction between two tensors of rank three with two indices contracted to create a tensor of rank two can be denoted as:
        $$
        {C_{\alpha}}^{\beta} = \sum_{\delta=1}^{N} \sum_{\gamma=1}^{M} {A_{\alpha \gamma}}^{\delta} {B^{\gamma \beta}}_{ \delta} \equiv {A_{\alpha \gamma}}^{\delta} {B^{\gamma \beta}}_{ \delta}
        $$
        The middle part shows how the contracted indices $\delta$ and $\gamma$ are summed over. The right-hand side instead follows what's known as "Einstein notation", which leaves out the summation signs. The indices that transfer to the output tensor are called \textit{free indices}, and the others are \textit{contracted indices}. In this example, the \textit{free indices} are $\alpha$ and $\beta$, and the \textit{contracted indices} are $\delta$ and $\gamma$. Contracted indies must have representations in both input tensors, meaning that both tensors must have indices of equal size that can be paired together. One of these indices must be "row mode" and the other "column mode", for those to be contracted indices. The free indices can appear in an arbitrary order in the output tensor. This means that the output can be presented in different permutations (arrangements of values). The number of free indices determines the number of possible permutations. Each value in the resulting tensor is calculated by calculating the product of each corresponding value in the sub-tensors that the \textit{contracted indices} create in the input tensors. These products are summed together to create the value in the output tensor. This means that in the example above, the value of $C$ at position $x$, $y$ is calculated as:
        $$
        {C_{x}}^{y} = \sum_{\delta=1}^{N} \sum_{\gamma=1}^{M} {A_{x \gamma}}^{\delta} {B^{\gamma y}}_{ \delta}
        $$
        \autoref{fig:part_tensor_contraction} visualizes the sub-tensors used to do this calculation. Note that both sub-tensors have the size $\gamma \times \delta$, meaning there is a corresponding value in each sub-tensor that is multiplied, and all multiplied values are summed to create the value at position $x$, $y$ in tensor $C$. This is very similar to how the dot product is calculated for vectors. In fact, a binary tensor contraction using two first-rank tensors where their indices are contracted is a dot product operation.
        
        In the field of computing science, there are seldom differences considered between superscript and subscript indices, and the notation is therefore only written with subscripts, making it look like this:
        $$
        {C_{\alpha}}_{\beta} = {A_{\alpha \gamma}}_{\delta} {B_{\gamma \beta}}_{ \delta}
        $$
        
    \subsection{BLAS}
    \label{sec:background:BLAS}
        BLAS(Basic Linear Algebra Subprograms) is an interface specification developed in FORTRAN for linear algebra operations \autocite{blas_1}. BLAS consists of three sets of operations. The first set of operations, "Level 1", was developed in 1979 and included scalar and vector operations. The second set of operations, "Level 2", was developed in 1988 and included matrix-vector operations \autocite{blas_2}. The third set of operations, "Level 3", was developed in 1990 and included matrix-matrix operations \autocite{blas_3}. The reasons for BLAS were that it would serve as a conceptual aid, improve the self-documentation quality of code, reduce execution time through optimized implementations, and avoid overlooking implementation subtleties \autocite{blas_1}. The BLAS specification includes reference implementations for its operations. These implementations are simple and slow, but they serve the purpose of exemplifying and showing how the interfaces are supposed to be used. High-performance usage of BLAS is achieved by linking with optimized BLAS implementations. Some open-source optimized implementations are ATLAS (Automatically Tuned Linear Algebra Software), which is the oldest one still in use \autocite{peise2017performance}, and GotoBLAS, which has been discontinued but whose content has been used and continued in OpenBLAS. Aside from open-source implementations, many hardware manufacturers have their own high-performance BLAS implementations. Some such manufacturers are Intel with oneMKL (oneAPI Math Kernel Library) \autocite{mkl}, Apple with Accelerate \autocite{accelerate}, Nvidia with cuBLAS \autocite{cublas}, IBM with ESSL (Engineering and Scientific Subroutine Library) \autocite{essl}, and AMD with AOCL-BLAS\autocite{aoclblas}, which is based on BLIS\autocite{blis}. BLIS is an open-source framework for implementing functionality that includes and extends BLAS. While BLAS covers basic operations, it has been used to form the basis for libraries with more advanced operations. The earliest such library is LINPACK \autocite{linpack}, which covers solvers for linear equations and least-squares problems. Another library called LAPACK (Linear Algebra PACKage) \autocite{lapack} has superseded LINPACK. Aside from the features of LINPACK, LAPACK also includes the features of EISPACK \autocite{eispack}, which is a collection of eigenvalue solvers. LAPACK is still under active development.
        \subsubsection{Interface}
            The BLAS specification covers many operations with its interfaces. These interfaces follow a convention where a prefix is used to specify the datatype used in the operation. These prefixes are s for single precision floating point numbers, d for double precision floating point numbers, c for single precision complex numbers, and z for double precision complex numbers \autocite{peise2017performance}. When no datatype is specified, there is a placeholder x instead.
            The interfaces for level 1 BLAS include basic vector operations such as dot product. The dot product interface looks like this:\\\\
            \texttt{xDOT(N, X, INCX, Y, INCY)}\\\\
            The inputs for the operations are the following:
            \begin{itemize}
                \item \textit{N} - The size of both vectors.
                \item \textit{X} - Pointer to the first value used in the first vector.
                \item \textit{INCX} - The increments for vector \textit{X}, meaning the memory jumps between values.
                \item \textit{Y} - Pointer to the first value used in the second vector.
                \item \textit{INCY} - The increments for vector \textit{Y}, meaning the memory jumps between values.
            \end{itemize}
            The dot product operation performs the dot product between vector \textit{X} and vector \textit{Y} and returns the results. For a mathematically correct dot product, one vector needs to be a row vector and the other a column vector, however, this interface does not differ between row and column vectors and therefore does not have that constraint.

            The interfaces for level 2 BLAS include matrix-vector operations such as matrix-vector multiplication. The matrix-vector multiplication interface looks like:\\\\
            \texttt{xGEMV(TRANS, M, N, ALPHA, A, LDA, X, INCX, BETA, Y, INCY)}\\\\
            The name stands for \textbf{general matrix-vector (multiplication)}. The inputs for the operations are the following:
            \begin{itemize}
                \item \textit{TRANS} - Decides if matrix \textit{A} will be transposed before the multiplication and how.
                \item \textit{M} - The number of rows of matrix \textit{A} and the size of vector \textit{Y}.
                \item \textit{N} - The number of columns of matrix \textit{A} and the size of vector \textit{X}.
                \item \textit{ALPHA} - A scalar for scaling the result of the multiplication.
                \item \textit{A} - Pointer to the first value in the matrix.
                \item \textit{LDA} - Leading dimensions of the matrix \textit{A}, meaning the size of the full matrix, which is bigger than the size of \textit{A} if \textit{A} is part of a larger matrix, and is the size of \textit{A} otherwise.
                \item \textit{X} - Pointer to the first value used in the input vector.
                \item \textit{INCX} - The increments for vector \textit{X}, meaning the memory jumps between values.
                \item \textit{BETA} - A scalar for scaling the vector \textit{Y} before the multiplication.
                \item \textit{Y} - Pointer to the first value used in the output vector.
                \item \textit{INCY} - The increments for vector \textit{Y}, meaning the memory jumps between values.
            \end{itemize}
            The operation transposes matrix \textit{A} depending on \textit{TRANS}. It can either not be transposed, transposed, or conjugate transposed. Further, it scales vector \textit{Y} by \textit{BETA}, performs matrix-vector multiplication between \textit{A} and \textit{X}, scales the result by \textit{ALPHA}, and adds it to \textit{Y}. Vector \textit{Y} contains the result of the operation. Mathematically, the operation looks like $y \leftarrow \alpha Ax + \beta y$, aside from how the matrix is transposed.
            The interfaces for level 2 BLAS include matrix operations such as matrix multiplication. The matrix multiplication interface looks like:\\\\
            \texttt{xGEMM(TRANSA, TRANSB, M, N, K, ALPHA, A, LDA, B, LDB, BETA, C, LDC)}\\\\
            The name stands for \textbf{general matrix-matrix (multiplication)}. The inputs for the operation are the following:
            \begin{itemize}
                \item \textit{TRANSA} - Decides if matrix \textit{A} will be transposed before the multiplication and how.
                \item \textit{TRANSB} - Decides if matrix \textit{B} will be transposed before the multiplication and how.
                \item \textit{M} - The number of rows of matrix \textit{A} and matrix \textit{C}.
                \item \textit{N} - The number of columns of matrix \textit{A} and the number of rows of matrix \textit{B}.
                \item \textit{K} - The number of columns of matrix \textit{B} and matrix \textit{C}.
                \item \textit{ALPHA} - A scalar for scaling the result of the multiplication.
                \item \textit{A} - Pointer to the first value in the first matrix.
                \item \textit{LDA} - Leading dimensions of the matrix \textit{A}, meaning the size of the full matrix, which is bigger than the size of \textit{A} if \textit{A} is part of a larger matrix and is the size of \textit{A} otherwise.
                \item \textit{B} - Pointer to the first value in the second matrix.
                \item \textit{LDB} - Leading dimensions of the matrix \textit{B}, meaning the size of the full matrix, which is bigger than the size of \textit{B} if \textit{B} is part of a larger matrix, and is the size of \textit{B} otherwise.
                \item \textit{BETA} - A scalar for scaling the matrix \textit{C} before the multiplication.
                \item \textit{C} - Pointer to the first value in the output matrix.
                \item \textit{LDC} - Leading dimensions of the matrix \textit{C}, meaning the size of the full matrix, which is bigger than the size of \textit{C} if \textit{C} is part of a larger matrix and is the size of \textit{C} otherwise.
            \end{itemize}
            The operation transposes matrix \textit{A} depending on \textit{TRANSA} and transposes matrix \textit{B} depending on \textit{TRANSB}. They can either not be transposed, transposed, or conjugate transposed. Further, matrix \textit{C} is scaled by \textit{BETA}, matrix \textit{A} and \textit{B} are multiplied, and the result is scaled by \textit{ALPHA} before being added to matrix \textit{C}. Matrix \textit{C} contains the result of the operation. Mathematically, the operation looks like $C \leftarrow \alpha AB + \beta C$, aside from the transpositions.

\section{Problem formulation}
\label{sec:problem_formulation}
    In the world of libraries for tensor operations, there are many packages that provide different sets of tensor operations. These packages are often independent of each other and, as a consequence, re-implement similar or equal functionality, often in a suboptimal way \autocite{psarras2021landscape}. BLAS has mitigated this for the world of vector and matrix operations by letting packages use the interfaces and then linking with optimized implementations. This same effect on the world of tensor operations would be needed and would decrease the replication of implementation efforts while also ensuring the usage of optimized implementations. The purpose of this paper is to work towards a similar standardization to BLAS, which would have the same desired effect on tensor operations. Creating a full standard covering all necessary operations is, however, a scope much larger than that for this master thesis. Therefore, this master thesis aims to only cover the operation of \textit{binary tensor contraction}. The question to answer in this master thesis is whether an interface in a BLAS-like style is sufficient to express an arbitrary \textit{binary tensor contraction} between two arbitrary tensors.
    
\section{Method}
\label{sec:method}
    This master thesis aims to discover whether an interface in a BLAS-like style is sufficient to express an arbitrary \textit{binary tensor contraction} between two arbitrary tensors. In this context, BLAS-like means the usage of a similar form to the interfaces specified in the BLAS specification, including the usage of primitive datatypes such as integers, floats, arrays, and so on, instead of structs or classes, making it portable. To answer this, the creation of such an interface has been attempted. BLAS was developed in the programming language FORTRAN, this interface was developed in the programming language C. The programming language C was used because, like FORTRAN, it is portable, meaning that it can easily be used by programs written in other programming languages \autocite{techopedia}. C is also nowadays more popular than FORTRAN \autocite{tiobe}. Just like BLAS, an example implementation will be developed alongside the interface. The example implementation will, together with tests, prove the correctness of the interface and serve as a reference for using the interface. The interface and implementation need to support the following features to fully support the operation: 
    \begin{itemize}
        \item The interface supports contractions between tensors of arbitrary dimensions.
        \item The interface supports contractions with an arbitrary number of contracted indices.
        \item The interface supports tensors that are sub-tensors of larger tensors.
        \item The reference implementation returns a correct result for all the previous cases.
    \end{itemize}
\section{Interface}
\label{sec:interface}
    The interface was developed through stages, where each stage added a new feature, starting with a simple matrix multiplication interface. By starting with a simple matrix multiplication interface, the development starts with a simplified version of an interface that BLAS supports, meaning that it is BLAS-like. By developing in stages by feature, the interface is kept concise and BLAS-like, containing only the necessary parameters. The name of the interface is xGETT, where \textit{GETT} stands for \textbf{general tensor tensor (contraction)}. The \textit{x} is a placeholder prefix. The prefix works just like BLAS, where it specifies the datatype used. For single precision floating point numbers, the prefix is \textit{s}, for double precision floating point numbers, the prefix is \textit{d}, for single precision complex numbers, the prefix is \textit{c}, and for double precision complex numbers, the prefix is \textit{z}. The arguments for the interface changed during the stages of development to be able to support new features.
    
    \subsection{Matrix multiplication}
    \label{sec:interface:matrix_multiplication}
    Starting with matrix multiplication. It is established that two matrices are used and that the width of the first matrix has to be the same size as the height of the second. Therefore, the only information that the interface needs is three numbers to represent the sizes of the matrices, and the memory addresses for the two input matrices and the output matrix. Taking inspiration from BLAS \texttt{xGEMM} interface described in \autoref{sec:background:BLAS}, the resulting interface looks as follows:
    \\\\
    \texttt{xGETT(M, N, K, A, B, C)}
    \\\\
    The inputs are:
    \begin{itemize}
        \item \textit{M} - The number of rows of matrix \textit{A} and matrix \textit{C}.
        \item \textit{N} - The number of columns of matrix \textit{A} and the number of rows of matrix \textit{B}.
        \item \textit{K} - The number of columns of matrix \textit{B} and matrix \textit{C}.
        \item \textit{A} - Pointer to the first used data point in one of the input matrices.
        \item \textit{B} - Pointer to the first used data point in the other input matrix.
        \item \textit{C} - Pointer to the first used data point in the output matrix.
    \end{itemize}
    \subsection{Non-contiguous storage}
    \label{sec:interface:non-contiguous_storage}
    Non-contiguous storage means that each memory section needed for the operation is not necessarily stored consecutively, there might be gaps. A smaller part of a bigger matrix might be used. For such memory storage, the interface needs to know the size of the full matrix, also known as the leading dimensions. If a smaller part of a bigger matrix is used, the leading dimensions would be the dimensions of the bigger matrix, otherwise, it would be the size of the used matrix. Again, taking inspiration from BLAS \texttt{xGEMM} interface, the resulting interface looks as follows:
    \\\\
    \texttt{xGETT(M, N, K, A, LDA, B, LDB, C, LDC)}
    \\\\
    The inputs are:
    \begin{itemize}
        \item \textit{M} - The number of rows of matrix \textit{A} and matrix \textit{C}.
        \item \textit{N} - The number of columns of matrix \textit{A} and the number of rows of matrix \textit{B}.
        \item \textit{K} - The number of columns of matrix \textit{B} and matrix \textit{C}.
        \item \textit{A} - Pointer to the first used data point in one of the input matrices.
        \item \textit{LDA} - Leading dimensions of matrix \textit{A}. An array of two values to determine the size of the two leading dimensions.
        \item \textit{B} - Pointer to the first used data point in the other input matrix.
        \item \textit{LDB} - Leading dimensions of matrix \textit{B}. An array of two values to determine the size of the two leading dimensions.
        \item \textit{C} - Pointer to the first used data point in the output matrix.
        \item \textit{LDC} - Leading dimensions of matrix \textit{C}. An array of two values to determine the size of the two leading dimensions.
    \end{itemize}
    \subsection{Selecting contracting dimension}
    \label{sec:interface:selecting_contracting_dimension}
    To support selecting which dimension to contract for the matrices, the interface needs to support matrices with sizes less dependent than earlier. It also needs to identify which dimension of each matrix to contract. This can easily be supported by adding an index input for each matrix. The index input would allow for indexing which matrix dimension to contract. The interface looks as follows:
    \\\\
    \texttt{xGETT(A, DIMA, EXTA, LDA, B, EXTB, CONTB, LDB, C, LDC)}
    \\\\
    The inputs are:
    \begin{itemize}
        \item \textit{A} - Pointer to the first used data point in one of the input matrices.
        \item \textit{EXTA} - The extents of matrix \textit{A}. An array of two values to determine the size of the two dimensions.
        \item \textit{CONTA} - Index deciding which of the dimensions of matrix \textit{A} to contract.
        \item \textit{LDA} - Leading dimensions of matrix \textit{A}. An array of two values to determine the size of the two leading dimensions.
        \item \textit{B} - Pointer to the first used data point in the other input matrix.
        \item \textit{EXTB} - The extents of matrix \textit{B}. An array of two values to determine the size of the two dimensions.
        \item \textit{CONTB} - Index deciding which of the dimensions of matrix \textit{B} to contract.
        \item \textit{LDB} - Leading dimensions of matrix \textit{B}. An array of two values to determine the size of the two leading dimensions.
        \item \textit{C} - Pointer to the first used data point in the output matrix.
        \item \textit{LDC} - Leading dimensions of matrix \textit{C}. An array of two values to determine the size of the two leading dimensions.
    \end{itemize}
    \subsection{Tensor multiplication}
    \label{sec:interface:tensor_multiplication}
    The interface does need both additional arguments and existing arguments require a higher degree of freedom to support using tensors instead of matrices. Firstly, inputs for the ranks of the input tensors are needed, the rank of the output tensor is possible to derive from the ranks of the input tensors and the number of contractions. Secondly, the arguments \textit{A}, \textit{B}, and \textit{C} still need to point to the first used data point, however, instead of only pointing to matrices, they can point to tensors of all ranks and sizes. Furthermore, \textit{EXTA} and \textit{EXTB} need to support extents for objects of a wider range of dimensions. \textit{LDA}, \textit{LDB}, and \textit{LDC} are renamed to \textit{INCA}, \textit{INCB}, and \textit{INCC} respectively. This will better represent their extended purpose. Instead of containing the dimensions of a possibly larger object, they contain the increments, or, in other words, the memory jumps between two data points in the object for all indices. This allows for sub-tensors of lower rank than the larger tensor they are a part of. The interface looks as follows:
    \\\\
    \texttt{xGETT(A, RANKA, EXTA, CONTA, INCA, B, RANKB, EXTB, CONTB, INCB, C, INCC)}
    \\\\
    The inputs are:
    \begin{itemize}
        \item \textit{A} - Pointer to the first used data point in one of the input tensors.
        \item \textit{RANKA} - The rank of tensor \textit{A}
        \item \textit{EXTA} - The extents of tensor \textit{A}. An array of values to determine the size of the dimensions.
        \item \textit{CONTA} - Index deciding which of the dimensions of tensor \textit{A} to contract.
        \item \textit{INCA} - Increments of tensor \textit{A}. An array that determines the memory jumps between two data points in each dimension.
        \item \textit{B} - Pointer to the first used data point in the other input tensor.
        \item \textit{RANKB} - The rank of tensor \textit{B}
        \item \textit{EXTB} - The extents of tensor \textit{B}. An array of values to determine the size of the dimensions.
        \item \textit{CONTB} - Index deciding which of the dimensions of tensor \textit{B} to contract.
        \item \textit{INCB} - Increments of tensor \textit{B}. An array that determines the memory jumps between two data points in each dimension.
        \item \textit{C} - Pointer to the first used data point in the output tensor.
        \item \textit{INCC} - Leading dimensions of tensor \textit{C}. An array of two values to determine the size of the two leading dimensions.
    \end{itemize}

    \subsection{Contracting multiple dimensions}
    \label{sec:interface:contracting_multiple_dimensions}
    The interface does both need an additional argument, and existing arguments require a higher degree of freedom to support the contraction of multiple dimensions. First, an argument for the number of contractions is needed. Secondly, the arguments \textit{CONTA} and \textit{CONTB} need to take an array of indices to contract instead of just one index. These arrays would be the length of the number of contractions. The interface looks as follows:
    \\\\
    \texttt{xGETT(A, RANKA, EXTA, CONTA, INCA, B, RANKB, EXTB, CONTB, INCB, C, INCC, CONTS)}
    \\\\
    The inputs are:
    \begin{itemize}
        \item \textit{A} - Pointer to the first used data point in one of the input tensors.
        \item \textit{RANKA} - The rank of tensor \textit{A}.
        \item \textit{EXTA} - The extents of tensor \textit{A}. An array of values to determine the size of the dimensions.
        \item \textit{CONTA} - An array of indices deciding which of the dimensions of tensor \textit{A} to contract.
        \item \textit{INCA} - Increments of tensor \textit{A}. An array that determines the memory jumps between two data points in each dimension.
        \item \textit{B} - Pointer to the first used data point in the other input tensor.
        \item \textit{RANKB} - The rank of tensor \textit{B}.
        \item \textit{EXTB} - The extents of tensor \textit{B}. An array of values to determine the size of the dimensions.
        \item \textit{CONTB} - An array of indices deciding which of the dimensions of tensor \textit{B} to contract.
        \item \textit{INCB} - Increments of tensor \textit{B}. An array that determines the memory jumps between two data points in each dimension.
        \item \textit{C} - Pointer to the first used data point in the output tensor.
        \item \textit{INCC} - Leading dimensions of tensor \textit{C}. An array of two values to determine the size of the two leading dimensions.
        \item \textit{CONTS} - The number of contractions.
    \end{itemize}
    
    \subsection{Selecting permutation}
    \label{sec:interface:selecting_permutation}
    The feature to choose a desired permutation \autocite{ttc} is less straightforward than other features. This is because to achieve it, an array is needed that indexes all the free indices of the two tensors and represents how these are placed in the resulting tensor. In this array, we also need to distinguish which index corresponds to which tensor. The following solutions have been considered:
    \begin{itemize}
        \item Indexing, where positive integers represent the first tensor and negative integers represent the second tensor. The position in the array would determine the placement of the free index in the resulting tensor. However, this indexing would need indexes to start from one and negative one because zero would overlap.
        \item Indexing, where the first tensor is indexed from zero and the second tensor is indexed from the number of indices of the first tensor. The position in the array would determine the placement of the free index in the resulting tensor.
        \item Indexing, where each index is a part of a pair, and the other part of the pair is an integer indexing which tensor it belongs to, zero for the first tensor and one for the second. The position in the array would determine the placement of the free index in the resulting tensor.
        \item Indexing, where each position represents a free index in one of the tensors, starting from the free indices in the first tensor and then the free indices in the second tensor. The integers written in each position of the array represent the placement of the indices in the resulting tensor.
    \end{itemize}
    The last option for the interface was chosen, making the interface:
    \\\\
    \texttt{xGETT(A, RANKA, EXTA, CONTA, INCA, B, RANKB, EXTB, CONTB, INCB, C, INCC, CONTS, PERM)}
    \\\\
    The inputs are:
    \begin{itemize}
        \item \textit{A} - Pointer to the first used data point in one of the input tensors.
        \item \textit{RANKA} - The rank of tensor \textit{A}.
        \item \textit{EXTA} - The extents of tensor \textit{A}. An array of values to determine the size of the dimensions.
        \item \textit{CONTA} - An array of indices deciding which of the dimensions of tensor \textit{A} to contract.
        \item \textit{INCA} - Increments of tensor \textit{A}. An array that determines the memory jumps between two data points in each dimension.
        \item \textit{B} - Pointer to the first used data point in the other input tensor.
        \item \textit{RANKB} - The rank of tensor \textit{B}.
        \item \textit{EXTB} - The extents of tensor \textit{B}. An array of values to determine the size of the dimensions.
        \item \textit{CONTB} - An array of indices deciding which of the dimensions of tensor \textit{B} to contract.
        \item \textit{INCB} - Increments of tensor \textit{B}. An array that determines the memory jumps between two data points in each dimension.
        \item \textit{C} - Pointer to the first used data point in the output tensor.
        \item \textit{INCC} - Leading dimensions of tensor \textit{C}. An array of two values to determine the size of the two leading dimensions.
        \item \textit{CONTS} - The number of contractions.
        \item \textit{PERM} - The permutation. An array that decides how the free indices are placed in the resulting tensor, deciding the permutation.
    \end{itemize}

    \subsection{Parameter order}
    \label{sec:interface:parameter_order}
    The ordering of the parameters should follow a logical pattern. The pattern chosen is that later parameters are described by their predecessors. Following that pattern, it becomes logical to first describe the input tensors. Starting with the tensor rank, followed by either the extents or the increments of the tensor, as both of them are independent of each other but dependent on the rank, and they both describe the tensor. The choice was made that the extent would be placed before the increments. After those comes the tensor pointer.

    Next are the parameters dictating how the operation will be performed. Starting with the number of contracted indices, followed by which indices are contracted for the first input tensor, followed by the same information for the second input tensor as the length of those arrays is the same as the number of contracted indices. After that comes the array dictating the order of the free indices because it is the length of the rank of both tensors subtracted by two times the number of contracted indices, and it is dependent on which indices are contracted. 

    The final section of the parameters describes the output tensor. Because the rank and extent can be deduced from the other parameters, the inputs for the output tensor are its increments and the pointer to the allocated data. Which is ordered with the increments first as they describe how the output tensor is distributed in memory.

    The final ordering of the interface is shown below:
    \\\\
    \texttt{xGETT(RANKA, EXTA, \hphantom{ }INCA, \hphantom{ }A,\\
    \null\qquad \quad RANKB, EXTB, \hphantom{ }INCB, \hphantom{ }B,\\
    \null\qquad \quad CONTS, CONTA, CONTB, PERM,\\
    \null\qquad \qquad \qquad \qquad \qquad \hphantom{ }INCC, \hphantom{ }C)}
    \begin{itemize}
        \item \textit{RANKA} - The rank of tensor \textit{A}.
        \item \textit{EXTA} - The extents of tensor \textit{A}. An array of values to determine the size of the dimensions.
        \item \textit{INCA} - Increments of tensor \textit{A}. An array that determines the memory jumps between two data points in each dimension.
        \item \textit{A} - Pointer to the first used data point in one of the input tensors.
        \item \textit{RANKB} - The rank of tensor \textit{B}.
        \item \textit{EXTB} - The extents of tensor \textit{B}. An array of values to determine the size of the dimensions.
        \item \textit{INCB} - Increments of tensor \textit{B}. An array that determines the memory jumps between two data points in each dimension.
        \item \textit{B} - Pointer to the first used data point in the other input tensor.
        \item \textit{CONTS} - The number of contractions.
        \item \textit{CONTA} - An array of indices deciding which of the dimensions of tensor \textit{A} to contract.
        \item \textit{CONTB} - An array of indices deciding which of the dimensions of tensor \textit{B} to contract.
        \item \textit{PERM} - The permutation. An array that decides how the free indices are placed in the resulting tensor, deciding the permutation.
        \item \textit{INCC} - Leading dimensions of tensor \textit{C}. An array of two values to determine the size of the two leading dimensions.
        \item \textit{C} - Pointer to the first used data point in the output tensor.
    \end{itemize}
    
\section{Implementation}
\label{sec:implementation}
    The reference implementation serves as a reference for how to use the interface to implement the operation. It also helps show the correctness of the interface because if the implementation works correctly, the interface can correctly support the operation.

\begin{figure}
        \centering
        \includegraphics[scale=0.35]{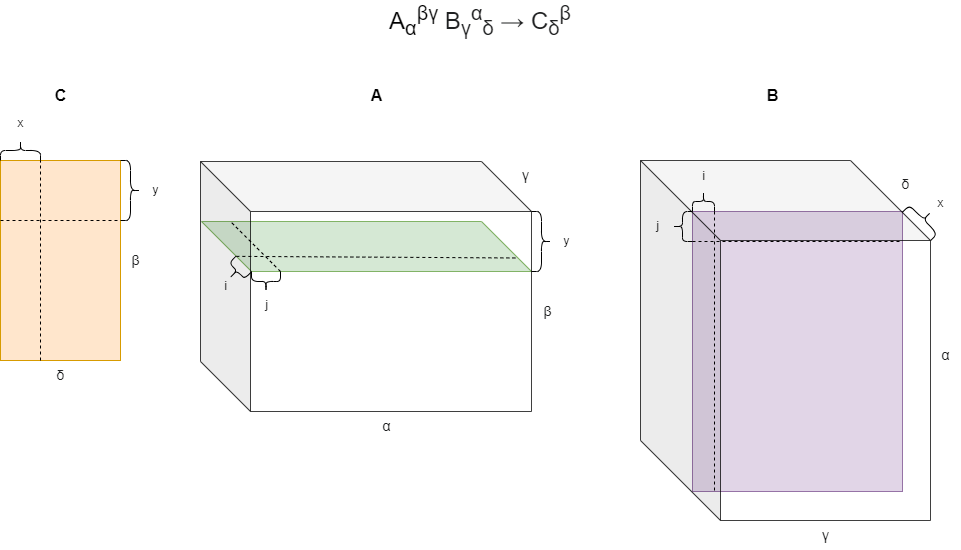}
        \caption{Display of the principles behind the workings of the implementation.}
        \label{fig:tensor_contraction}
    \end{figure}

    The implementation starts by processing the input data in a manageable format. If we call the input tensors A and B and the output tensor C. The data needed is firstly the size of C(the size that the free indices make up)(the amount of data points), an array of the increments of C, and an array of the extents of C(the extent of the free indices) ordered according to the desired permutation dictated by the \textit{PERM} input. More data needed are an array keeping track of which extent of C is inherited from which tensor among A and B, an array of the increments of the input tensors for the free indices(ordered like the extents of C), the size of contracted indices(the amount of data points to contract from each of the tensors), and the number of contracted indices. Lastly, an array of the extents of the contracted indices is needed, as are two arrays containing the increments of the contracted indices for the input tensors(ordered according to the order of the extents of the contracted indices). Aside from the data, two additional arrays are needed. One has the same size as the number of free indices and keeps track of which part of C is manipulated and which parts of A and B are contracted, coordinates in the free indices. The other has the same size as the number of contracted indices and keeps track of which part of the contracted area of A and B is in use,  coordinates in the contracted indices. These arrays are incremented throughout the calculation and then follow the extents, meaning that the first value is the first to be incremented until it reaches its corresponding extent, then it resets, and the next value is incremented, and so on for the rest of the values. \autoref{fig:tensor_contraction} visualizes a contraction between two three-dimensional tensors, creating a two-dimensional tensor. In that case, the first array would contain the values $x$ and $y$, while the second array would contain $i$ and $j$.
    
    When the data is in a manageable format, the calculation of the values for the output tensors begins. The algorithm loops through each position of tensor C. The position incrimination is achieved by using the extent of C and the array that keeps track of coordinates in the free indices. The increments of C are used to calculate the correct position in memory. As shown in \autoref{fig:tensor_contraction}, the position in tensor C corresponds to areas in tensors A and B. These areas could be seen as sub-tensors for the corresponding tensor, and both are equally sized and have corresponding values. For each position in tensor C, the corresponding sub-tensors in both A and B are looped through, where each product between the sub-tensors' corresponding values is summed together, creating the value for C at that position. The position incrimination is achieved by using the extent of the contracted indices and the array that keeps track of coordinates in the contracted indices. The increments for the free indices for tensors A and B and the increments for their contracted indices are used to calculate the correct positions in memory. 
    
\begin{figure}
        \centering
        \includegraphics[scale=0.5]{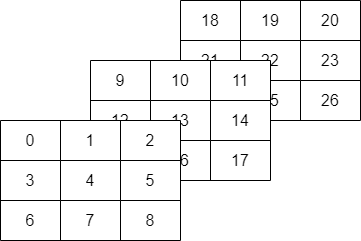}
        \caption{A rank three tensor indexed in memory with increments [1, 3, 9].}
        \label{fig:tensor_indexing}
    \end{figure}

\begin{figure}
        \centering
        \includegraphics[scale=0.075]{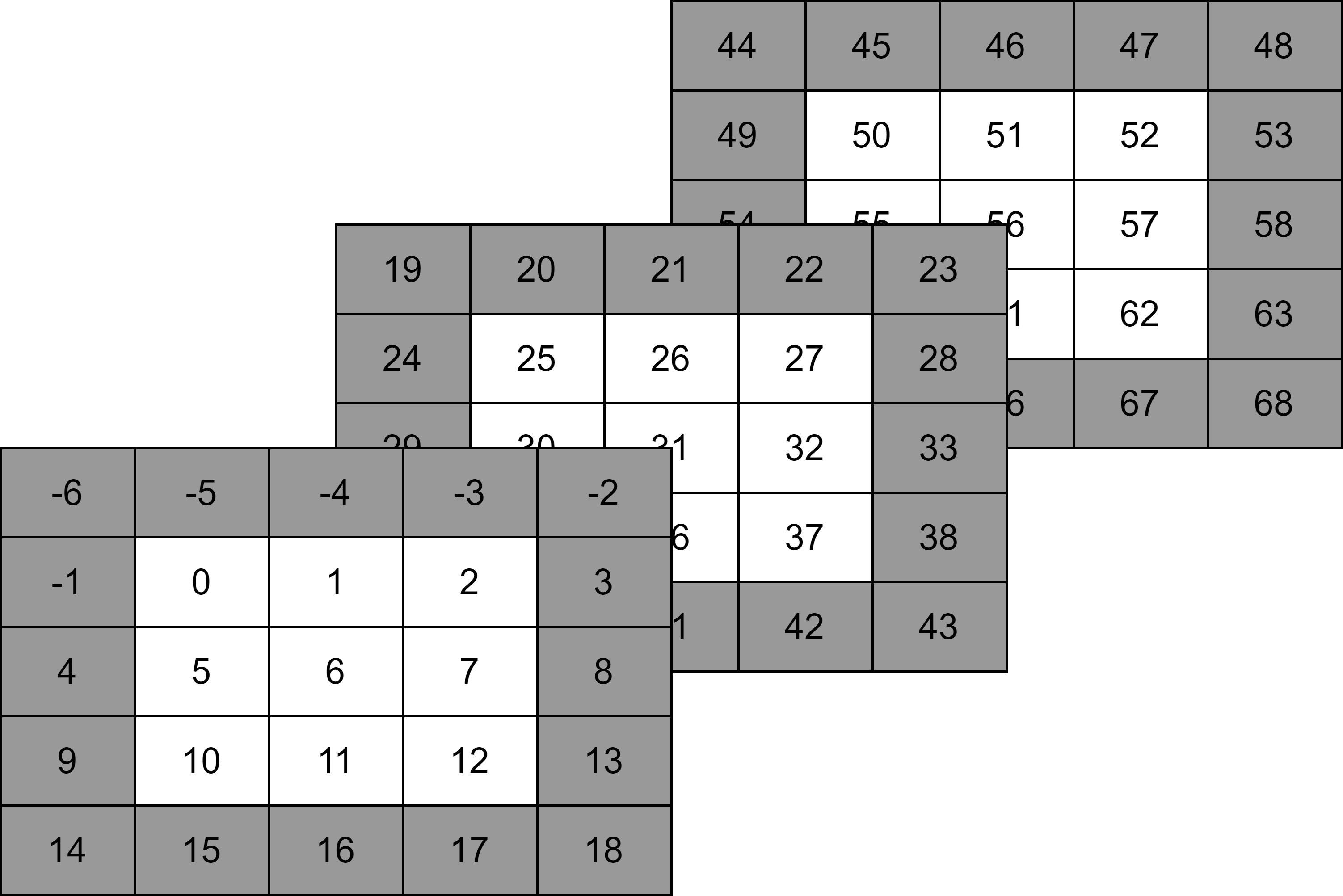}
        \caption{A rank three sub-tensor indexed in memory with increments [1, 5, 25]. The gray parts are outside the used sub-tensor but are parts of the bigger tensor.}
        \label{fig:sub-tensor_indexing}
    \end{figure}

\begin{figure}
        \centering
        \includegraphics[scale=0.5]{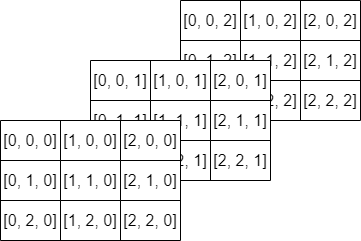}
        \caption{Coordinates for the data in a rank three tensor.}
        \label{fig:tensor_coordinates}
    \end{figure}

    \autoref{alg:tensor_contraction} describes in pseudocode how the algorithm works. The code consists of the following variables:
    \begin{itemize}
        \item \textit{SizeFreeInd} - the size of free indices
        \item \textit{SizeContInd} - the size of contracted indices
        \item \textit{NumFreeInd} - number of free indices
        \item \textit{NumContInd} - number of contracted indices
        \item \textit{FreeIdxA} - indexing tensor A using only free indices
        \item \textit{FreeIdxB} - indexing tensor B using only free indices
        \item \textit{IdxA} - indexing in tensor A
        \item \textit{IdxB} - indexing in tensor B
        \item \textit{IdxC} - indexing in tensor C
        \item \textit{TenFreeInd} - an array keeping track of which free indices belong to which input tensor, either A or B
        \item \textit{CoordFreeInd} - an array keeping track of the coordinates for the free indices
        \item \textit{CoordContInd} - an array keeping track of the coordinates for the contracted indices
        \item \textit{ExtFreeInd} - an array of the extents of free indices
        \item \textit{ExtContInd} - an array of the extents of contracted indices
        \item \textit{IncA} - increments of tensor A
        \item \textit{IncB} - increments of tensor B
        \item \textit{IncC} - increments of tensor C
        \item \textit{IncFreeInd} - increments of free indices
        \item \textit{A} - tensor A
        \item \textit{B} - tensor B
        \item \textit{C} - tensor C
    \end{itemize}
    In \autoref{alg:tensor_contraction}, rows 5-12 are responsible for calculating the index for the current position in tensor C and the corresponding position in tensors A and B. An integer is calculated for each of the tensors representing the index for the position represented by $x$ and $y$ in \autoref{fig:tensor_contraction}. This is done by multiplying coordinates with increments for all free indices and summing them. \autoref{fig:tensor_indexing} shows how a non-sub-tensor could be indexed in memory, \autoref{fig:sub-tensor_indexing} shows how a sub-tensor could be indexed in memory, and \autoref{fig:tensor_coordinates} shows how the coordinates for both tensors would be distributed. Rows 13-22 go through all positions in the contracted area of tensors A and B and perform the contraction at row 20. Rows 16-19 calculate the indexing of the contracted area. The indexing calculation is done in the same way as for the free indices by multiplying increments with coordinates for all contracted indices and summing them together with the indexing of the free indices. \autoref{alg:increment_coordinates} describes how the function \textit{IncCoords} works. 

    \begin{algorithm}
        \caption{Binary Tensor Contraction} 
        \begin{algorithmic}[1]        
            \For{$i \leftarrow 0$ to $SizeFreeInd - 1$}
                \State $IdxC \leftarrow 0$
                \State $FreeIdxA \leftarrow 0$
                \State $FreeIdxB \leftarrow 0$
                \For{$j \leftarrow 0$ to $NumFreeInd - 1$}
                    \State $IdxC \leftarrow IdxC + IncC[j] * CoordFreeInd[j]$
                    \If{$TenFreeInd[j] = A$}
                        \State $FreeIdxA \leftarrow FreeIdxA + IncFreeInd[j] * CoordFreeInd[j]$
                    \Else
                        \State $FreeIdxB \leftarrow FreeIdxB + IncFreeInd[j] * CoordFreeInd[j]$
                    \EndIf
                \EndFor
                \For{$j \leftarrow  0$ to $SizeContInd - 1$}
                    \State $IdxA \leftarrow FreeIdxA$
                    \State $IdxB \leftarrow FreeIdxB$
                    \For{$k \leftarrow  0$ to $NumContInd - 1$}
                        \State $IdxA \leftarrow IdxA + IncA[k] * CoordContInd[k]$
                        \State $IdxB \leftarrow IdxB + IncB[k] * CoordContInd[k]$
                    \EndFor
                    \State $C[IdxC] \leftarrow C[IdxC] + A[IdxA] * B[IdxB]$
                    \State \Call{$IncCoords$}{$NumContInd, CoordContInd, ExtFreeInd$} 
                \EndFor
                \State \Call{$IncCoords$}{$NumFreeInd, CoordFreeInd, ExtContInd$}
            \EndFor
        \end{algorithmic} 
        \label{alg:tensor_contraction}
    \end{algorithm}

    \begin{minipage}{\textwidth}
    \autoref{alg:increment_coordinates} uses the variables:
    \begin{itemize}
        \item \textit{NumOfInd} - the number of indices
        \item \textit{Coords} - an array the size of the number of indices keeping coordinates for a value. This will be incremented in the function.
        \item \textit{Exts} - the extents for the indices
        \item \textit{i} - indexing among indices
    \end{itemize}
    \end{minipage}
    
    The algorithm works by incrementing coordinates corresponding to indices and taking the modulo between the incremented coordinate and its corresponding extent. This means that if a coordinate becomes as large as an extent, it loops back to zero. The algorithm starts by incrementing the first coordinate. If it becomes zero after the modulo, the algorithm continues with the next coordinate. As long as the modulo of the last incremented coordinate becomes zero, the algorithm continues with the next until all coordinates have been incremented. For \autoref{fig:tensor_coordinates} this would mean [0, 0, 0], [1, 0, 0], [2, 0, 0], [0, 1, 0], [1, 1, 0], [2, 1, 0], and so on, following the indexing shown in \autoref{fig:tensor_indexing}.
    \begin{algorithm}
        \caption{Increment Coordinates} 
        \begin{algorithmic}[1]
            \Procedure{$IncCoords$}{$NumOfInd, Coords, Exts$}
                \If{$NumOfInd \leq 0$}
                    \State \Return
                \EndIf

                \State $i\leftarrow0$
                \Do
                    \State $Coords[i] \leftarrow (Coords[i] + 1) \% Exts[i]$
                    \State $i \leftarrow i + 1$
                \DoWhile{$Coords[i - 1] = 0$ \& $i < NumOfInd$}
            \EndProcedure
        \end{algorithmic} 
        \label{alg:increment_coordinates}
    \end{algorithm}
\section{Testing}
\label{sec:testing}
    The interface, along with its example implementation, has been tested against two libraries with similar functionality. The first one is TBLIS \autocite{tblis}, a library and framework for performing tensor operations. TBLIS is developed in C++ but has a C interface. TBLIS has an operation called \textit{tensor multiplication} that has similar capabilities to the interface and implementation developed in this master thesis but is more abstract. This operation can directly be used to test the interface and implementation developed in this master thesis. The second library is Tensor Toolbox \autocite{tensor_toolbox}, which is a MATLAB library for tensor operations. To make it possible to call the functions of Tensor Toolbox through C code, MATLAB Engine was used. MATLAB Engine is a tool that makes it possible to call MATLAB code from other programming languages. Tensor Toolbox does not contain a single function that includes all the functionality of the interface and implementation developed in this thesis. It does not support sub-tensors, which is compensated for by giving the functions tensors that are extracted from sub-tensors. The library contains a \textit{tensor times tensor} operation that does binary tensor contractions but does not give the option to choose permutation. However, this can be compensated for by using the \textit{permute} operation that the library contains.

    The tests that are needed are basic operation tests such as simple contractions, commutativity(meaning that the order of the matrices does not matter), and correct ordering of indices. Identified edge cases are when no indices are contracted, contraction to a scalar, contracting with a scalar, contracting with a vector, and contractions where all extents of all indices of all tensors are the same. Furthermore, tests are needed for non-contiguous storage.

    For the test, code has been written that generates tensors for a possible binary tensor contraction. This generation is somewhat random, it always results in correct contractions, and a set of parameters determines the traits of the tensors that are generated to be contracted. Minimum and maximum rank for each tensor, minimum and maximum extents for each tensor, and how many indices are to be contracted are some of these parameters.

    The following tests have been performed on the implemented operation:
    \begin{itemize}
        \item \textbf{Basic contraction} - an arbitrary contraction between two tensors with ranks between one and five that are not sub-tensors, with between zero and four contracted indices.
        \item \textbf{Commutativity} - an operation is commutative if the order of the input does not affect the outcome. To test this, two arbitrary contractions between two tensors with ranks between one and five and which are not sub-tensors, with between zero and four contracted indices, where the second contraction uses the same input tensors but swaps their places in the interface. It also needs to rearrange the order of indices for the output tensor, as that order is dependent on the tensors' order.
        \item \textbf{Nothing contraction} - a contraction between two tensors with ranks between one and five that are not sub-tensors, with zero contracted indices.
        \item \textbf{Scalar contraction} - a contraction between two tensors with ranks between one and five that are not sub-tensors, with all indices contracted, resulting in a scalar.
        \item \textbf{Permutations} - testing if the ordering of indices works correctly. The test creates an arbitrary contraction between two tensors with ranks between one and five that are not sub-tensors. This contraction is done $n$ times, where $n$ is the rank of the output tensor. For each contraction, the order of indices is shifted to the left.
        \item \textbf{Rank zero tensor} - a contraction between a tensor with ranks between one and five and a tensor with rank zero(scalar), none of them are sub-tensors. No indices can be contracted.
        \item \textbf{Rank one tensor} - a contraction between a tensor with ranks between one and five and a tensor with rank one(vector). Between one and zero indices are contracted.
        \item \textbf{Square tensors zero contractions} - a contraction between two tensors of rank two with all extents the same, between one and five, no tensors are sub-tensors. No indices contracted.
        \item \textbf{Square tensors one contraction} - a contraction between two tensors of rank two with all extents the same, between one and five, no tensors are sub-tensors. One index contracted.
        \item \textbf{Square tensors two contractions} - a contraction between two tensors of rank two with all extents the same, between one and five, no tensors are sub-tensors. Two indices contracted.
        \item \textbf{Cube tensors zero contractions} - a contraction between two tensors of rank three with all extents the same, between one and five, no tensors are sub-tensors. No indices contracted.
        \item \textbf{Cube tensor one contraction} - a contraction between two tensors of rank three with all extents the same, between one and five, no tensors are sub-tensors. One index contracted.
        \item \textbf{Cube tensor two contractions} - a contraction between two tensors of rank three with all extents the same, between one and five, no tensors are sub-tensors. Two indices contracted.
        \item \textbf{Cube tensor three contractions} - a contraction between two tensors of rank three with all extents the same, between one and five, no tensors are sub-tensors. Three indices contracted.
        \item \textbf{Hypercube tensors zero contractions} - a contraction between two tensors of rank four with all extents the same, between one and five, no tensors are sub-tensors. No indices contracted.
        \item \textbf{Hypercube tensor one contraction} - a contraction between two tensors of rank four with all extents the same, between one and five, no tensors are sub-tensors. One index contracted.
        \item \textbf{Hypercube tensor two contractions} - a contraction between two tensors of rank four with all extents the same, between one and five, no tensors are sub-tensors. Two indices contracted.
        \item \textbf{Hypercube tensor three contractions} - a contraction between two tensors of rank four with all extents the same, between one and five, no tensors are sub-tensors. Three indices contracted.
        \item \textbf{Hypercube tensor four contractions} - a contraction between two tensors of rank four with all extents the same, between one and five, no tensors are sub-tensors. Four indices contracted.
        \item \textbf{Sub-tensor of same rank} - an arbitrary contraction between two tensors with ranks between one and five. All tensors, including the output tensor, are sub-tensors randomly placed in a larger tensor that have the same rank, but the extents are between one and five data points larger. Between zero and four contracted indices.
        \item \textbf{Negative increment} - an arbitrary contraction between two tensors with ranks between one and five that are not sub-tensors, with between zero and four contracted indices. Instead of all increments being positive, they are negative, meaning that the tensor is technically read backward in memory.
        \item \textbf{Sub-tensor negative increment} - an arbitrary contraction between two tensors with ranks between one and five. All tensors, including the output tensor, are sub-tensors randomly placed in a larger tensor whose extents are between one and five data points larger. Between zero and four contracted indices. Instead of all increments being positive, they are negative, meaning that the tensor is technically read backward in memory.
        \item \textbf{Sub-tensor of lower rank} - an arbitrary contraction between two tensors with ranks between one and five, and between zero and four contracted indices. All tensors, including the output tensor, are sub-tensors randomly placed in a larger tensor whose rank is between one and three higher and extents between one and three data points larger.
    \end{itemize}

\section{Conclusion}
\label{sec:conclusion}
    The interface 
    \\\\
    \texttt{xGETT(RANKA, EXTA, \hphantom{ }INCA, \hphantom{ }A,\\
    \null\qquad \quad RANKB, EXTB, \hphantom{ }INCB, \hphantom{ }B,\\
    \null\qquad \quad CONTS, CONTA, CONTB, PERM,\\
    \null\qquad \qquad \qquad \qquad \qquad \hphantom{ }INCC, \hphantom{ }C)}
    \\\\
    that has been developed follows a BLAS-like style. It also fulfills the earlier-mentioned requirements for handling the binary tensor contraction operation:
    \begin{itemize}
        \item The interface supports contractions between tensors of arbitrary dimensions.
        \item The interface supports contractions with an arbitrary number of contracted indices.
        \item The interface supports tensors that are sub-tensors of larger tensors.
        \item The reference implementation returns a correct result for all the previous cases.
    \end{itemize}
    This is proven by the implementation and the tests performed. This means that an interface in a BLAS-like style is sufficient to express an arbitrary binary tensor contraction between two arbitrary tensors. This interface, however, is not the only possibility for a BLAS-like interface for binary tensor contractions. The amount of data that the interface requires is the least amount found possible in this thesis. Another interface could rearrange or combine inputs for a smaller interface. However, it is arguable if that would benefit the interface.

\section{Future Work}
\label{sec:future_work}
    This project is limited to only the binary tensor contraction because of time constraints. However, a full specification like BLAS, but for tensor operations requires interfaces for more operations. This work shows that a BLAS-like interface is possible for binary tensor contractions, and thus the possibility of similar interfaces for other tensor operations being possible is high. Especially as binary tensor contraction is one of the more complex operations. The interface also lays a foundation for how interfaces for other operations might look and how they can be developed. Some other operations that would be needed in a full interface specification for tensor operations are:
    \begin{itemize}
        \item Reduction
        \item Transpose
        \item Set to scalar
        \item Trace
        \item Addition
        \item Hadamard product
        \item Weighting
        \item Scale
        \item Shift
    \end{itemize}
\newpage

\printbibliography

\end{document}